\documentclass[11pt, reqno, english]{smfart}

\usepackage[T1]{fontenc}
\usepackage[english, francais]{babel}

\usepackage{amsmath}
\usepackage{amssymb}
\usepackage{url,xspace,smfthm}

\usepackage{epsfig}
\usepackage{graphicx}
\graphicspath{{figs/}}

\makeatletter
    \def\ps@copyright{\ps@empty
    \def\@oddfoot{\hfil\small\copyright 2015, \SMF}}
\makeatother

\usepackage[svgnames, table]{xcolor}
\definecolor{oneblue}{rgb}{0.0, 0.0, 0.85}

\usepackage[colorlinks,
            urlcolor=oneblue,
            linkcolor=LightSlateGrey,
            citecolor=gray,
            bookmarksopen=false,
            pagebackref]{hyperref}
\hypersetup{pdfpagemode=UseNone}

\addtocontents{toc}{\setcounter{tocdepth}{2}}

\newcommand{\ie}{\emph{i.e.}~}
\newcommand{\eg}{\emph{e.g.}~}
\newcommand{\etal}{\emph{et al.}~}

\newcommand{\SMF}{R.~Escobedo, D.~Dutykh, \etal}
\newcommand{\BibTeX}{{\scshape Bib}\kern-.08em\TeX}
\newcommand{\T}{\S\kern .15em\relax }
\newcommand{\AMS}{$\mathcal{A}$\kern-.1667em\lower.5ex\hbox
        {$\mathcal{M}$}\kern-.125em$\mathcal{S}$}

\newenvironment{myepigraph}
  {\par\hfill\itshape
   \begin{tabular}{@{}r@{\hspace{2em}}}} 
  {\end{tabular}\par\medskip}

\title{Group Size Effect on the Success of Wolves Hunting}
\date{June 24, 2015}

\author[R.~Escobedo]{Ram\'on Escobedo}
\thanks{Corresponding author: R.~Escobedo.}
\address{BCAM -- Basque Center for Applied Mathematics, Alda. Mazarredo 14, 48009 Bilbao, Spain \and AEPA-Euskadi, Pte. de Deusto 7, 48013 Bilbao, Spain}
\email{escobedor@gmail.com}
\urladdr{http://www.aepa-euskadi.org/escobedo/}

\author[D.~Dutykh]{Denys Dutykh}
\address{CNRS -- LAMA UMR 5127, Universit\'e Savoie Mont Blanc, Campus Scientifique, 73376 Le Bourget-du-Lac, France}
\email{Denys.Dutykh@univ-savoie.fr}
\urladdr{http://www.denys-dutykh.com/}

\author[C.~Muro]{Cristina Muro}
\address{AEPA-Euskadi, Pte. de Deusto 7, 48013 Bilbao, Spain}
\urladdr{http://www.aepa-euskadi.org/}

\author[L.~Spector]{Lee Spector}
\address{School of Cognitive Science, Hampshire College, Amherst, MA 01002, USA}
\urladdr{http://faculty.hampshire.edu/lspector/}
\email{lspector@hampshire.edu}

\author[R.P.~Coppinger]{Raymond P. Coppinger}
\address{School of Cognitive Science, Hampshire College, Amherst, MA 01002, USA}
\urladdr{https://www.hampshire.edu/faculty/raymond-coppinger/}

\subjclass{37N25; 92-08; 92B05}
\keywords{Collective animal behavior; social foraging in carnivores; computational agent-based model; dynamical systems; stability and bifurcation}

\altkeywords{Comportement animalier collectif;}

\begin{document}
\def\smfbyname{}

\frontmatter


\begin{abstract}
Social foraging shows unexpected features such as the existence of a group size threshold to accomplish a successful hunt. Above this threshold, additional individuals do not increase the probability of capturing the prey. Recent direct observations of wolves ({\it Canis lupus}) in Yellowstone Park show that the group size threshold when hunting its most formidable prey, bison ({\it Bison bison}), is nearly three times greater than when hunting elk ({\it Cervus elaphus}), a~prey that is considerably less challenging to capture than bison. These observations provide empirical support to a computational particle model of group hunting which was previously shown to be effective in explaining why hunting success peaks at apparently small pack sizes when hunting elk. The model is based on considering two critical distances between wolves and prey: the minimal safe distance at which wolves stand from the prey, and the avoidance distance at which wolves move away from each other when they approach the prey. The minimal safe distance is longer when the prey is more dangerous to hunt. We show that the model explains effectively that the group size threshold is greater when the minimal safe distance is longer. Actually, the model reveals that the group size threshold results from the nonlinear combination of the variations of both critical distances. Although both distances are longer when the prey is more dangerous, they contribute oppositely to the value of the group size threshold: the group size threshold is smaller when the avoidance distance is longer. This unexpected mechanism gives rise to a global increase of the group size threshold when considering bison with respect to elk, but other prey more dangerous than elk can lead to specific critical distances that can give rise to the same group size threshold. Our results show that the computational model can guide further research on group size effects, suggesting that more experimental observations should be obtained for other kind of prey as {\it e.g.}~moose ({\it Alces alces}).
\end{abstract}

\begin{altabstract}
La recherche collective de la nourriture montre des ph\'enom\`enes inattendus comme l'existence de la taille optimale pour un groupe de chasseur afin de r\'eussir la chasse. Au-dessus de ce seuil, les individus suppl\'ementaires n'augmentent pas la probabilit\'e de capturer la proie. Des observations directes r\'ecentes des loups ({\it Canis lupus}) dans Yellowstone Park montrent que ce seuil pour la chasse d'un bison ({\it Bison bison}) est environ trois fois plus grand que dans la chasse d'un cerf \'elaphe ({\it Cervus elaphus}), une proie qui est consid\'erablement plus simple \`a capturer. Ces observations fournissent des donn\'ees empiriques pour un mod\`ele math\'ematique qui d\'ecrit le processus de la chasse d'un groupe de loups. Ce mod\`ele a \'et\'e valid\'e r\'ecemment et ses pr\'edictions semblent indiquer que la taux de succ\`es a un maximum pour les groupes de taille plut\^ot mod\'er\'ee. Ce mod\`ele fait intervenir deux distances critiques: la distance minimale de s\'ecurit\'e entre les loups et la proie, et une autre distance d'\'evitement (cette fois-ci entre les loups) afin d'assurer la s\'ecurit\'e lorsqu'ils s'approchent de la proie. La distance minimale de s\'ecurit\'e est d'autant plus grande que la proie est dangereuse. Dans cette \'etude nous montrons que le seuil pour la taille de groupe est plus grand lorsque la distance de s\'ecurit\'e est plus longue. En effet, le mod\`ele montre que l'existence de ce seuil provient d'une combinaison nonlin\'eaire complexe des deux distances de s\'ecurit\'e \`a la fois. Bien que ces deux distances sont plus longues lorsque la proie est plus dangereuse, elles contribuent de mani\`ere oppos\'ee \`a la valeur du seuil: la taille critique du groupe est plus petite lorsque la distance d'\'evitement est plus longue. Ce m\'ecanisme inattendu a pour l'effet l'augmentation globale du seuil lorsqu'on consid\`ere un bison par raport \`a un cerf mais les autres proies plus dangeureuses que le cerf peuvent avoir des distances de s\'ecurit\'e telles que le seuil serait le m\^eme que pour le bison. Le bon accord entre nos r\'esultats et les observations montrent que ce mod\`ele math\'ematique peut \^etre utilis\'e afin d'\'etudier les effets sur la taille du groupe optimal. D'autres observations sont n\'ecessaires sur les autres types de la proie comme un \'elan ({\it Alces alces}), par exemple.
\end{altabstract}


\maketitle

\vspace{-0.5em}
\tableofcontents
\newpage

\mainmatter

\vspace{.8cm}
\begin{myepigraph}
``Plus on est de fous, plus on rit''.\\
({\it The more the merrier})\\[.5ex]
La maison de campagne,\\
Dancourt (1688).
\end{myepigraph}

\section{Introduction}

Applied to social foraging, this French proverb illustrates the intuitive idea that the greater the number of individuals participating in a hunt, the easier the capture of the prey. It explains also a second idea: the greater the number of hunters, the larger the prey they can capture.

Whatever the social circumstances, whoever uses this allocution is always conscious of the incontrovertible fact that {\it there is a limit}. In fact, observational data from a range of large social predators show that above an optimal group size, the benefit per individual participating in the hunt does not increase and can even decline; see~\cite{MacNulty2012} and the extensive list of references therein. Moreover, this optimal group size is surprisingly small, ranging from 2 to 5 in carnivores~\cite{MacNulty2012}, which leads to the hypothesis that there is probably not support among the fundamental evolutionary forces for living in groups.

There are two potential reasons for the leveling of hunting success in groups whose sizes are above the optimal value: 
\begin{enumerate}
  \item interference between inept hunters,
  \item individual withholding of effort.
\end{enumerate}
Experimental and theoretical research is currently exploring which one of the two factors has the most important contribution.

Recently, MacNulty {\it et~al.}~(2012) \cite{MacNulty2012} reported and analyzed wolf ({\it Canis lupus}) observational data when hunting elk ({\it Cervus elaphus}) in Yellowstone National Park, finding that, for wolf-pack sizes greater than $N = 4$, where $N$ is the number of wolves participating in the hunt, a decline in wolf effort is responsible for impeding large groups from reaching a greater hunting success: wolves withhold effort to reduce high hunting costs such as injury (the rate with which a wolf's performance decreases is correlated with the danger associated with the task)~\cite{MacNulty2012}.

Escobedo {\it et~al.}~(2014) \cite{Escobedo2014} used a computational particle model to evaluate the existence of a physical mechanism by which complex behavioral patterns emerge in groups greater than the optimal size observed in nature~\cite{MacNulty2012}. These complex patterns result from destabilization of a regular polygonal formation that wolf-packs tend to adopt when surrounding a prey. The radius of this regular polygon varies in time according to the instantaneous value of a critical distance, namely, the {\it safe distance $d_c(t)$}, at which wolves position themselves so as not to be injured by the prey. The distances for elk, bison and moose are shown in Fig.~1. The peril of being injured increases as distance decreases~\cite{MacNulty2009}, so that when wolves arrive at this safe distance $d_c(t)$ they cease to approach the prey. The distance $d_c(t)$ varies during the hunt due to variation in how this peril is perceived at each instant; as the prey gets tired, $d_c(t)$ decreases, but can rise abruptly if the prey prompts a sudden counterattack. When $d_c(t)$ is smaller than a critical threshold $d_c^*$, the polygonal formation loses its stability and the stable spatial configuration becomes multi-orbital: the wolves are distributed along (at least) two orbits, with one orbit closer to the prey and one or more orbits further from the prey than the vertices of the (now unstable) polygonal formation.

Escobedo {\it et~al.}~(2014) \cite{Escobedo2014} hypothesizes that the multi-orbital configuration induces the emergence of privileged positions, and therefore of disadvantageous positions, and that this leads to the disruption of the hunt~\cite{Escobedo2014}. They showed that the threshold $d_c^*$ is greater in larger pack sizes, so that $d_c(t)$ takes values under $d_c^*$ more easily in larger packs, so that the hunt is more easily disrupted in larger packs; see Fig.~1 in~\cite{Escobedo2014}. Thus, an optimal pack size exists above which~$d_c(t)$ decreases below $d_c^*$ systematically, therefore compromising the hunting success.

Very recently, MacNulty {\it et~al.}~(2014) \cite{MacNulty2014} have reported observational data about wolves hunting their most formidable prey, bison ({\it Bison bison}), again in Yellowstone, where bison are three times more difficult to kill (by wolves) than elk~\cite{MacNulty2014}. The main observation is that, again, there exists an optimal wolf-pack size at which hunting success levels off, and that this wolf-pack size is $N = 11$, a fairly common (and not so small) wolf-pack size (which can reach 25 individuals~\cite{Mech2003}). MacNulty {\it et al.} \cite{MacNulty2014} attribute the increase in optimal group size to two possible factors:
\begin{enumerate}
  \item a higher level of cooperation between wolves when hunting larger prey, due to the very low capture rate of a single hunter,
  \item the stabilization of the spatial configurations displayed by large packs, due to the observation that the safe distance between the wolves and the prey is longer when facing more dangerous prey.
\end{enumerate}
Noticeably, MacNulty~{\it et~al.}~(2014) \cite{MacNulty2014} base this second hypothesis on the extrapolation of the insights provided by the computational particle model of Escobedo {\it et~al.}~(2014) \cite{Escobedo2014}.

The present work is thus motivated by, first, the new observational data presented by MacNulty {\it et~al.}~(2014) \cite{MacNulty2014}, and second, by the explicit mention of our model as one of the two potential explanations of the observed phenomenon. We show here that the interpretation of MacNulty~{\it et al.}~(2014) \cite{MacNulty2014} corroborates the modelled conclusions that, for more dangerous prey, the threshold of the safe distance under which the regular pack formation is unstable, $d_c^*$, is greater. However, the model reveals that the mechanism of variation of the optimal pack size for hunting success is more complex and unpredictable, as it results from the nonlinear combination of two effects with opposite contribution to the variation of the optimal pack size. We present this mechanism in the next section {\it Hypothesis}; our results and our discussion are presented together in section {\it Results}, and we conclude in section {\it Conclusions}. Materials and methods are presented in the supporting information.


\subsection{Hypothesis}

Our hypothesis is that the way a wolf approaches a prey is different from one prey species to another. The prudence with which a wolf moves near a prey defines two characteristic safe distances to the prey: the {\it avoidance distance} $d_a$, at which wolves move away from each other to have a better vision of the prey and enough room for escaping maneuvers; and the {\it minimal safe distance} $d_s$, shorter than $d_a$, at which wolves move away from the prey not to be harmed by the horns or the legs of the prey. We assume that all wolves have the same perception of danger, so that each prey species defines specific values of $d_s$ and $d_a$, characterized by the morphological and behavioral traits of each prey species. Thus, both $d_s$ and $d_a$ are assumed to remain constant during the hunt.

\begin{figure}
\centerline{
\epsfxsize=0.99\textwidth
 \epsfbox{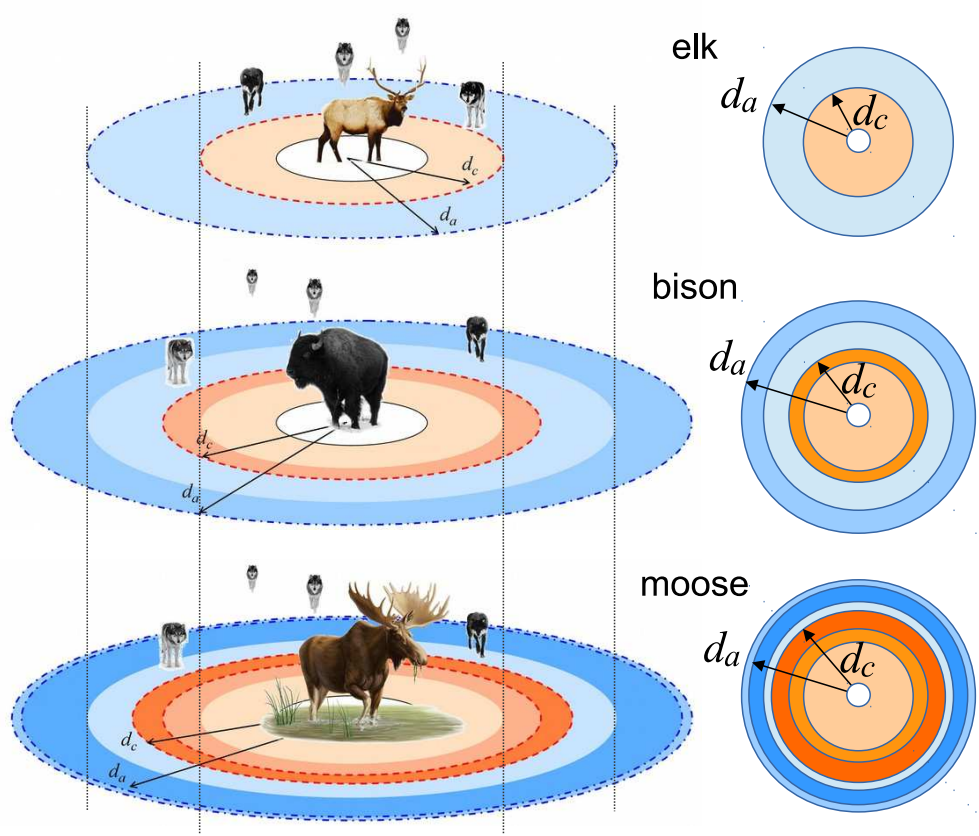}
}
\vspace{0.2cm}
\caption{\small\em Safe distances $d_s$ and $d_a$ for (A) elk, (B) bison and (C) moose. Dashed line: minimal safe distance $d_s$ at which a wolf would stand from the prey; dot-dashed line: avoidance distance $d_a$ at which wolves move away from each other when they approach the prey. Both $d_s$ and $d_a$ are assumed greater when facing bison.}
\label{foto}
\end{figure}

The hypothesis is that the more dangerous the prey, the greater the values of $d_s$ and $d_a$, but not necessarily in the same proportion: a prey can be more dangerous, but only at closer distances, if, {\it e.g.}, it is much weightier, or only at longer distances, if, {\it e.g.}, the horns are larger. Similarly, when close to the end, some prey trigger sudden counterattacks towards individual wolves, while some others sweep the area around them with their horns. Fig.~\ref{foto} illustrates this diversity showing the three circular regions around the prey defined by $d_s$ and $d_a$, whose width depends on the prey species, for three typical prey of wolf-packs: elk, bison and moose. The general formulation of our hypothesis is as follows. Let $S_1$ be a prey species less dangerous than $S_2$. Then, we have
\begin{equation*}
  (i) \; d_s^2 = \alpha_s d_s^1, \qquad 
  (ii)\; d_a^2 = \alpha_a d_a^1, \qquad \mbox{with } \alpha_{s,a}>1.
\end{equation*}
This means that $d_s^2 > d_s^1$ and $d_a^2 > d_a^1$, but not necessarily in the same proportion, \ie, $\alpha_s$ is not necessarily of the same order than $\alpha_a$.

In the particular case of bison and elk, bison are much larger than elk, more aggressive, and more likely to injure or kill wolves that attack them~\cite{MacNulty2014}, so
\begin{equation*}
  (i) \; d_s^{\rm bison} > d_s^{\rm elk}, \qquad (ii)\;  d_a^{\rm bison} > d_a^{\rm elk},
\end{equation*}
and we assume that $\alpha_s \approx \alpha_a$, although this is not necessarily the case for other species.

The avoidance distance $d_a$ was already introduced in~\cite{Escobedo2014}; $d_s$ is introduced here for the first time in the context of this model, although the instantaneous safe distance $d_c(t)$ was also already used in~\cite{Escobedo2014} (but was introduced in~\cite{Muro2011}). As we showed in~\cite{Escobedo2014}, the bifurcation threshold $d_c^*$ is determined univocally by the pack size $N$ and the avoidance distance $d_a$: $d_c^* \equiv d_c^*(N,d_a)$. Thus, the minimal safe distance $d_s$, which is a kind of lower bound of $d_c(t)$ for all $t>0$, can be smaller or greater than $d_c^*$.

In fact, it is precisely this relation that will determine the optimal pack size. The argument is as follows. During a hunt, the instantaneous safe distance $d_c(t)$ is always greater than the absolute minimal safe distance $d_s$. If $d_c^*$ is smaller than $d_s$, then $d_c(t)$ can not reach the bifurcation value $d_c^*$ at which complex patterns emerge, because $d_c^*<d_s<d_c(t)$, so the hunt is never disrupted, meaning that $N$ is smaller or equal to the optimal pack size $N_{\rm OPT}(d_s,d_a)$. However, if $d_c^*$ is greater than $d_s$, then $d_c(t)$ can take values below $d_c^*$ and complex patterns can emerge, meaning that, for the given value of $d_a$, the pack size $N$ is greater than the optimal value $N_{\rm OPT}(d_s,d_a)$.

The study thus reduces to obtain computationally the value of $d_c^*$ as a function of $N$ and $d_a$, and then to determine the optimal size for a given value of $d_s$. This is carried out in the next Section {\it Results}, finding that, for a fixed $d_a$, $N_{\rm OPT}(d_s,d_a)$ is greater when $d_s$ is greater, and that, for a fixed $d_s$, $N_{\rm OPT}(d_s,d_a)$ is smaller when $d_a$ is greater.


\section{Results}

\begin{figure}
\epsfxsize=0.99\textwidth
\begin{center}
 \epsfbox{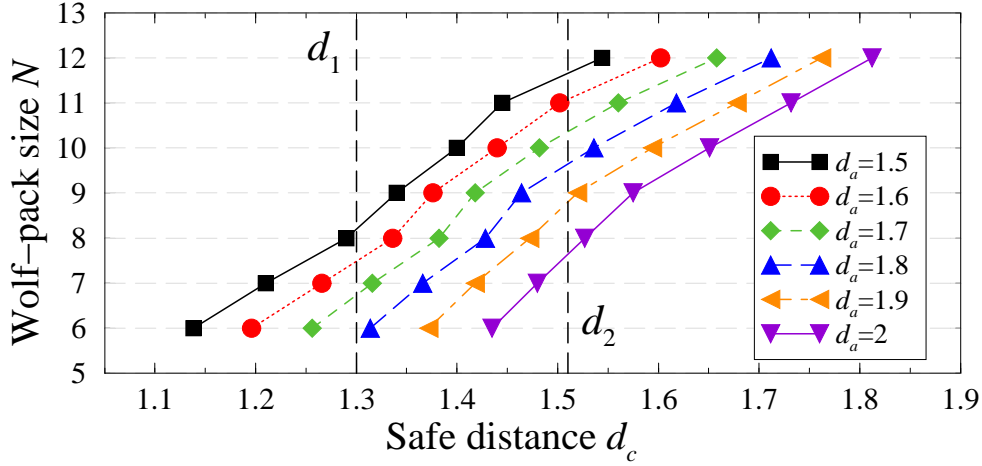}
\end{center}
\caption{\small\em Bifurcation points $d_c^*(N,d_a)$ for different values of $N$ and $d_a$. Symbols denote the bifurcation points. Lines joining symbols corresponds to the same value of $d_a$. First line starting from the left is $d_a^{\rm elk} = 1.5$, detailed in Fig.~\ref{fig2}. Successive lines to the right correspond to increments of $\Delta d_a = 0.1$.}
\label{fig1}
\end{figure}

We used the computational particle model introduced in~\cite{Escobedo2014} to derive the safe distance threshold $d_c^*(N,d_a)$ as a function of $N$ and $d_a$ for $N = 2,\dots,12$ and $d_a= 1.5$ to 2 with increments of $\Delta d_a = 0.1$. The numerical method is described in Section {\it Materials and Methods}. The result is shown in Fig.~\ref{fig1}.

\begin{figure}
\epsfxsize=0.99\textwidth
\begin{center}
 \epsfbox{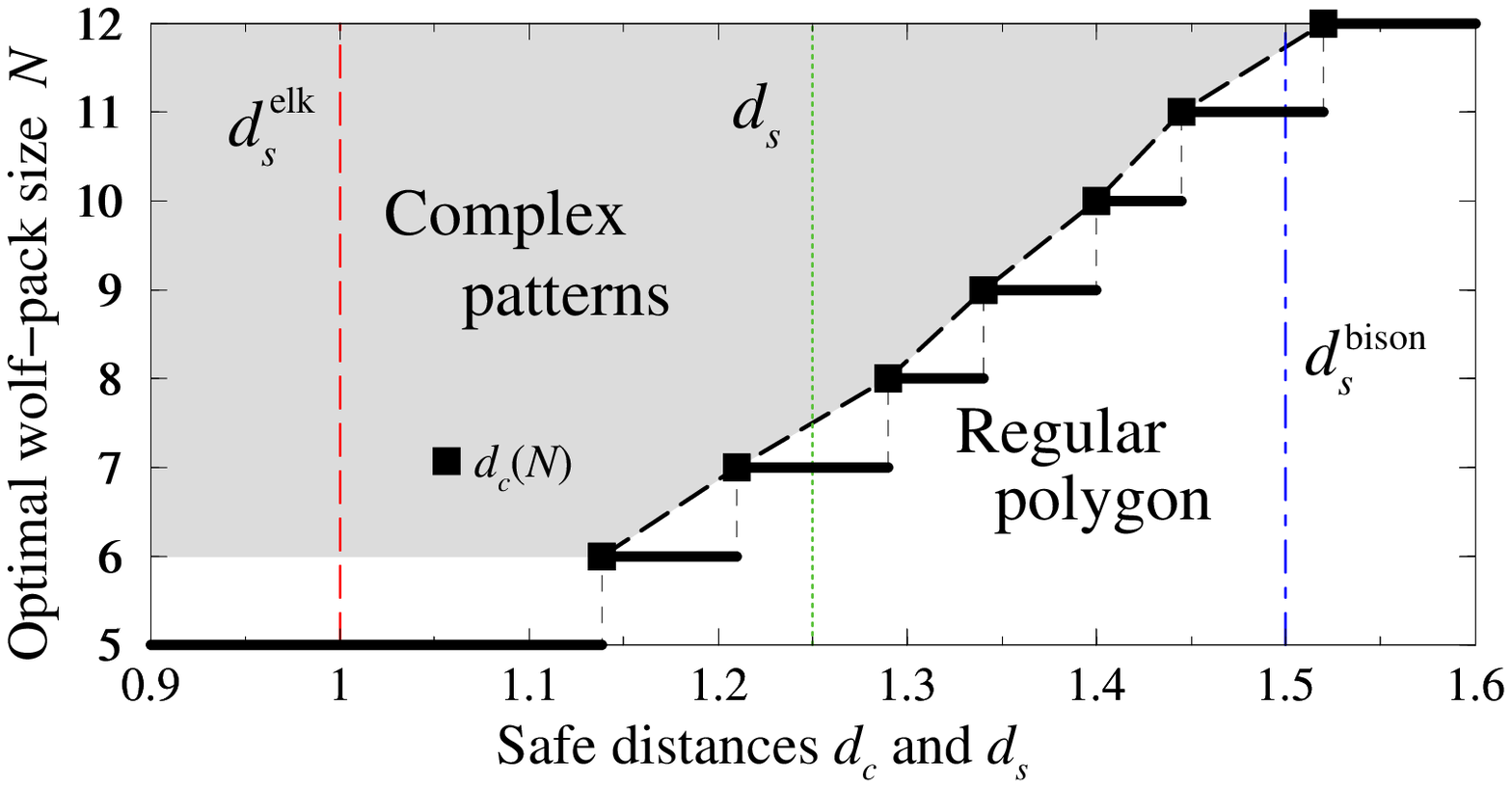}
\end{center}
\caption{\small\em Optimal wolf-pack size $N_{\rm OPT}(d_s,d_a)$ for a fixed value of $d_a = 1.5$. Filled squares denote the bifurcation points $d_c^*(N,d_a)$ delimiting the region where complex behavioral patterns emerge (gray region). Horizontal wide segments denote the optimal wolf-pack size for $d_s \in [0.9,1.6]$. Vertical lines denote the minimal safe distance $d_s$ for two kinds of prey, $d_s^{\rm elk} = 1$ (dashed line) and $d_s^{\rm bison} = 1.5$ (dot-dashed line), and a third example for $d_s = 1.25$ (dotted line). See also Fig.~1 in~\cite{Escobedo2014}.}
\label{fig2}
\end{figure}

Let us analyze first the particular case $d_a = d_a^{\rm elk}=1.5$, whose data were obtained in~\cite{Escobedo2014}. The rest of data are presented here for the first time. See then Fig.~\ref{fig2}.

For $N \le 5$, no bifurcation points exist: the regular polygon is stable for all $d_c(t)$). For $N > 5$, there is always a bifurcation point $d_c^*(N,d_a)$ that separates the interval of values of $d_c(t)$ where complex behavioral patterns emerge ($d_c < d_c^*$) from the interval of values for which the wolf-pack tends to the regular polygonal formation ($d_c^* < d_c$).

The time-varying safe distance $d_c(t)$ can decrease below $d_c^*$ only if $d_s < d_c^*$. Then, if~$d_s>d_c^*(N)$, the optimal group size for hunting success is greater or equal to $N$, and if $d_s < d_c^*(N)$, the optimal group size is smaller than $N$. Thus, if $d_c^*(N) < d_s < d_c^*(N$+$1)$ for $N > 5$, the optimal group size is $N$ because $d_c(t)$ can take values below $d_c^*(N+1)$ (and complex formations of packs of $N+1$ wolves), but not below $d_c^*(N)$ (so that the packs of $N$ wolves will always display the regular polygon of $N$ vertices).

This result does not depend on the value of $d_a$, so that the general result can be formulated as follows: the optimal wolf-pack size $N_{\rm OPT}(d_s, d_a)$ is determined by
\begin{equation}\label{e1}
  N_{\rm OPT}(d_s, d_a) = \max\big\{N\ge 5: d_c^*(N,d_a) < d_s < d_c^*(N + 1,d_a) \big\},
\end{equation}
where we have considered that $d_c^*(N,d_a) = 0$ for $N\le5$ and for all $d_a$. See the wide horizontal segments in Fig.~\ref{fig2}, denoting the value of $N_{\rm OPT}$ as a function of $d_s$ for the fixed value of $d_a = 1.5$. This result means that the value $N_{\rm OPT}(d_s^{\rm elk},d_a^{\rm elk})=5$ we obtained in~\cite{Escobedo2014} when hunting elk is valid provided $d_s^{\rm elk}<d_c^*(6) \approx 1.14$. Another example is shown in Fig.~\ref{fig2} for $d_s = 1.25$: $d_c^*(7) \approx 1.21$, $d_c^*(8) \approx 1.29$, so $d_c(t)$ can take values in $[d_s,d_c^*(8)]$ but not in $[d_c^*(7),d_s]$, so the optimal size is $N_{\rm OPT} = 7$.

Expression \eqref{e1} shows that, for a fixed value of $d_a$, the optimal pack size $N_{\rm OPT}(d_s,\cdot)$ is an increasing function of $d_s$. On the other hand, Fig.~\ref{fig1} shows that for greater values of $d_a$, the lines of bifurcation points appear as if they were shifted to the right, and this, quite regularly: an increment of $\Delta d_a = 0.1$ produces a shift of $\Delta d_c \approx 0.06$ to the right.\footnote{An empirical relation is $d_c^*(N,d_a) = d_c^*(N,d_a^{\rm elk}$ + $\Delta d_a) = d_c^*(N, d_a^{\rm elk}) + k \Delta d_c$, for $k=0,\dots,5$.} Thus, the region of complex patterns propagates to the right as $d_a$ grows, meaning that, for a fixed value of $d_s$, $N_{\rm OPT}(\cdot,d_a)$ is a decreasing function of $d_a$.

Let us show an example. Assume that $d_s = 1.4$ and $d_a = 1.7$. For a wolf-pack of size $N = 8$, we obtain $d_c^*(N$ = $8, d_a$ = $1.7) \approx 1.38$, so the point $(d_s, N) = (1.4, 8)$ is in the white region, $d_c^* < d_s$, and wolf-packs of size $N = 8$ will never reach the threshold distance under which complex patterns are triggered. In turn, for a wolf-pack of size $N = 9$, the point $(d_s,N) = (1.4, 9)$ is in the Grey region, so that $d_c(t)$ can take values below $d_c^*(9)$, thus triggering complex patters and disrupting the hunt. Therefore, the optimal wolf-pack size for $(d_s,d_a) = (1.4, 1.7)$ is $N = 8$.

Assume now that $d_a = 1.8$. Then, the Grey region expands to the right, $d_c^*(N$ = $8, d_a$ = $1.7) < d_c^*(N$ = $8, d_a$ = $1.8)$, meaning that $d_c(t)$ can more easily decrease below $d_c^*$. Then, wolf-packs of size $N=8$ will display complex patterns when $d_c(t) \in (d_s, d_c^*)$, so that the optimal wolf-pack size for  $(d_s,d_a) = (1.4, 1.8)$ is $N < 8$.

The conclusion is that both distances $d_s$ and $d_a$ contribute oppositely to the variation of the optimal wolf-pack size when the prey is more dangerous:
\begin{equation}
  d_s \mbox{ is larger } \Rightarrow N_{\rm OPT} \mbox{ is larger};\quad
  d_a \mbox{ is larger } \Rightarrow N_{\rm OPT} \mbox{ is smaller}.
\end{equation}
The variation of the optimal pack size when the prey is more dangerous is therefore the result of the nonlinear combination of two opposite effects: a longer safe distance at which wolves can stand from the prey enables more individuals to occupy a single (stable) orbit, but a longer avoidance distance requires more space (for escaping maneuvers from a more dangerous prey), such that the single orbit is destabilized, and the wolf-pack is split into a multi-orbital configuration.

Such a nonlinear combination gives rise generally to a modulated increase of the optimal pack size when the kind of prey changes from ordinary to more dangerous. This is what happens to the model when we change from elk to bison. In this case, the respective increments of $d_s$ and $d_a$ are assumed to be of the same order. This leads to a net increase from $N_{\rm OPT}^{\rm elk} = 5$ (for, say, $d_s^{\rm elk} = 1.1$ and $d_a^{\rm elk} = 1.5$, as in~\cite{Escobedo2014}), to $N_{\rm OPT}^{\rm bison} = 11$ (as in~\cite{MacNulty2014}, and that our model reproduces, {\it e.g.}~for $d_s^{\rm bison} = 1.7$ and $d_a^{\rm bison} = 1.9$). However, a decrease of the optimal pack size can also take place, provided the increment in $d_a$ is greater than the increment in $d_s$.

Let us illustrate this nonlinear effect with an example. Consider a prey $S_1$ with $(d_s^{(1)}, d_a^{(1)}) = (1.3, 1.5)$. Then, $N^{(1)}_{\rm OPT}(1.3, 1.5) = 8$. Now consider a more dangerous prey $S_2$ with $(d_s^{(2)}, d_a^{(2)}) = (1.51,1.6)$, {\it i.e.}, $d_s^{(2)}$ is quite larger than $d_s^{(1)}$ but $d_a^{(2)}$ is slightly larger than $d_a^{(1)}$. Then, the optimal pack size grows to $N^{(2)}_{\rm OPT}(1.51,1.6) = 11$. However, if $d_a^{(2)}$ is also quite larger than $d_a^{(1)}$, then $N^{(2)}_{\rm OPT}$ can remain unchanged ($N^{(2)}_{\rm OPT}(1.51,1.9) = 8$ for $d_a^{(2)} = 1.9$), or even decrease ($N^{(2)}_{\rm OPT}(1.51,2) = 7$ for $d_a^{(2)} = 2$).

This probably could happened for other prey species (\eg, moose) or in special hunting conditions (\eg, snow), where large horns and large legs are an advantage to repel wolves. Our results serve also for other prey species, for which the variation of the optimal wolf-pack size with respect to the case of the elk is not necessarily straightforward.


\section{Conclusion}

We have illustrated that MacNulty {\it et al.}~(2014) made a good prediction based on our model. We have presented a mathematical formulation of the hypothesis about the main factor making hunting success to level off at an optimal group size also when hunting formidable prey, which is the disruption of the group for sizes greater than this optimal group size. We have confirmed that the optimal group size is generally larger when hunting more dangerous prey. Moreover, we have revealed an unexpected nonlinear mechanism which contributes to modulate this increase of the optimal size. The mechanism consists of the nonlinear combination of two opposite effects induced respectively by the increase of both critical distances $d_s$ and $d_a$ when the prey is more dangerous. Our results show that the model is able to explain the recently reported observational data, thus validating our hypothesis. The model will serve to guide researchers in further observations, in particular to consider other prey such as moose ({\it Alces alces}, see Fig.~\ref{foto}) and other ungulates. Of special interest would be real data about critical distances.


\section*{Acknowledgments}

We thank Karyn Coppinger and Lorna Coppinger. R.E.~and D.D.~acknowledges the support of the Spanish association of assistance dogs AEPA-Euskadi, the Severo Ochoa Program SEV-2013-0323 of the MINECO and the Basque Government BERC 2014-2017 Program. R.E.~is supported by the Advanced Grant FP7-246775 NUMERIWAVES of the European Research Council Executive Agency, FA9550-14-1-0214 of the EOARD-AFOSR, FA9550-15-1-0027 of AFOSR and the MTM2011-29306-C02-00 Grant of the MINECO. This material is based upon work supported by the National Science Foundation under Grants No.~1017817, 1129139, and 1331283. Any opinions, findings, and conclusions or recommendations expressed in this publication are those of the authors and do not necessarily reflect the views of the National Science Foundation.


\section{Methods and model}

We use a computational particle model where particles represent wolves and prey and their behavior obeys Newton's second law. Interactions between agents (wolf-prey and wolf-wolf) are described in terms of radial attractive and repulsive forces. Critical distances define regions where interactions change qualitatively. We use continuous interaction functions so that transitions between regions are smooth. The model describes the dynamics of a pack of $N$ wolves hunting a single prey. The description covers the phase of the hunt which starts right after the phase where a prey has been isolated from the group of prey and is pursued by the wolves, and is valid until the killing phase starts and the prey stops moving definitively, it falls down, and it is dissected by the wolves.

We consider that the hunt consists of a series of fast and slow dynamic transitions of the prey-wolves system between stable spatial configurations determined by the time-variation of the critical distance $d_c(t)$. Depending on the duration of these transitions with respect to the time required to enter the basin of attraction of the stable states, the system may adopt a spatial configuration where wolves are uniformly distributed in one single orbit around the prey or may exhibit abrupt changes between multi-orbital configurations. At the beginning of the hunt, the prey displays an energetic behavior with abrupt changes in its trajectory and can even face one or more wolves. There, $d_c(t)$ is at its maximum value. As time evolves, the prey becomes tired and wolves gain confidence to get closer to the prey, so that $d_c(t)$ decreases. Before falling down, the prey may exhibit sudden counterattacks, making $d_c(t)$ to rise abruptly. The prey can find a way to escape, or, alternatively, repeat the strategy, making $d_c(t)$ to variate smoothly or abruptly, until the prey falls down or escapes definitively.

The model is described in detail in~\cite{Escobedo2014}, where numerical simulations of the spatial configurations and the convergence process of the system towards the stable configuration are provided for a number of cases. The emphasis in~\cite{Escobedo2014} was to describe in detail how complex patterns emerge and lead to the disruption of the hunt due to the interaction of wolves, both in the multi-orbital configurations and in the transition between stable configurations when $d_c$ is close to $d_c^*$. We obtained the stable spatial configurations towards which the prey-wolves system converges when $d_c$ is kept constant in time for different values of the pack size $N$. We also established the parameter conditions under which the cohesion of the system is preserved, that is, the distance from wolves to the prey is bounded so that wolves do not go to infinity. The study was carried out for the case of a middle size prey (elk), and the size of a pack was considered constant during the hunt.

The characterization of the stable configurations is done in a reduced model where the effects of noise and perturbations have been removed, in order to expose the essential features of the model which are responsible for the specific patterns under study. Adding noise and perturbation do not qualitatively change the results, that is, for a given value of $(N, d_c)$, the perturbed model gives rise to a configuration qualitatively identical to the stable configuration given by the unperturbed model for this value of $(N, d_c)$. Quantitative differences can slightly affect, for example, the critical value of $d_c^*$, the exact location of the agents (but not the geometric formation of the pack), the orientation of the flocking motion, or the velocity of rotation of the milling formation.

Here we focus on the effect of a larger prey (bison), and we show that the same model serves to illustrate that more dangerous prey allow a greater number of wolves to participate in the hunt before the critical pack size at which hunting success peaks is reached. This is done by simply considering a minimum safe distance $d_s$ as the closest distance to the prey that wolves can reach; $d_s$ is the distance that wolves will never cross, unless they enter in the killing phase, where the prey is already down. The distance $d_s$ is of course larger if the prey is larger or is perceived as more dangerous by the wolves; see Fig.~\ref{foto}. Considering the optimal size as a function of the two critical distances, \ie~$N_{\rm OPT} \equiv N_{\rm OPT}(d_s,d_a)$, the first argument is thus that $d_s^{\rm elk} < d_s^{\rm bison}$ implies $N_{\rm OPT}^{\rm elk} < N_{\rm OPT}^{\rm bison}$, that is, the optimal size is an increasing function of $d_s$.

The identification of the second mechanism that contributes to the variation of the threshold of the group size when the prey is perceived as more dangerous is based on the fact that the critical distance $d_a$ is also larger when facing a more dangerous prey. Numerical simulations provide us with the critical value $d_c^*$ for different values of $d_a$ (see Fig.~\ref{fig1}), thus illustrating that this mechanism contributes to the decrease of the optimal wolf-pack size. The second argument is thus that $d_a^{\rm elk} < d_a^{\rm bison}$ implies $N_{\rm OPT}^{\rm elk} > N_{\rm OPT}^{\rm bison}$, that is, the optimal size is a decreasing function of $d_s$. It is the nonlinear combination of these two effects that gives rise to the resulting optimal wolf-pack size.


\subsection{The model}

The prey is denoted by $P$ and the wolves by $W_i$, $i = 1,\dots,N$, where $N$ is the wolf-pack size. The position of the agents (prey and wolves) is denoted by the $N+1$ vectors $\vec{u}_i(t) = (x_i(t), y_i(t))$, $i = p,1,\dots,N$. Wolves and prey obey Newton's second law $m \vec{a} = \vec{\mathcal{F}}$, where $\vec{a} = \dot{\vec{v}}$ is the acceleration vector, $\vec{v} = \dot{\vec{u}}$ is the velocity vector and $\vec{\mathcal{F}}$ is the resultant of the forces acting on the agent. The dot denotes derivation with respect to time ($\vec{a} = \dot{\vec{v}} =\ddot{\vec{u}}$).

The dynamical system consists of $2(N+1)$ ordinary differential equations,
\begin{eqnarray*}
  \dot{\vec{u}}_i(t)    & = & \vec{v}_i(t),\quad i = p, 1, \dots,N,\\
  m_p\dot{\vec{v}}_p(t) & = & \sum_{i = 1}^N \vec{F}_{p,i}(t) - \nu_p \vec{v}_p(t),\\
  m_i\dot{\vec{v}}_i(t) & = & \vec{F}_{i,p}(t) + \sum_{j = 1,j\ne i}^N \vec{F}_{i,j}(t) - \nu_i \vec{v}_i(t), \quad i = 1,\dots,N,
\end{eqnarray*}
where $\vec{F}_{p,i}$, $i = 1,\dots,N$, are the $N$ repulsive forces exerted by the wolves on the prey, $\vec{F}_{i,p}$, $i=1,\dots,N$, is the long-range attractive and short-range repulsive force $\vec{F}_{i,p}$ exerted by the prey on the $i$-th wolf, $\vec{F}_{i,j}$, $j = 1,\dots,N$, $j \ne i$, is the repulsive forces exerted on th e$i$-th wolf by the other $N-1$ wolves, and $-\nu_i \vec{v}_i$, $i = p, 1, \dots, N$, are ground friction forces in the opposite direction of motion, with coefficient of friction $\nu$, considered identical for all wolves, $\nu_i = \nu_j$, $\forall i, j = 1,\dots,N$, and larger for preys than for wolves, $\nu_p > \nu_i$.

Attractive-repulsive interaction forces between two agents are described by radial functions based on the distance separating both agents. Here we use the classical formulation of Gazi \& Passino~\cite{Gazi2004}. Other interaction potentials (Lennard-Jones, Morse) can be used. More precisely, we use the specific one introduced by Shi \& Xie~\cite{Shi2011} and interpreted in the original model~\cite{Escobedo2014} as the most biologically realistic, because there the repulsion increases to infinity as the distance between two individuals goes to zero, and the attraction decreases to zero as the distance grows to infinity.

More explicitly, the system can be written as follows,
\begin{eqnarray}
  \dot{\vec{u}}_i(t) & = & \vec{v}_i(t), \quad i = p, 1, \dots, N, \\
  \dot{\vec{v}}_p(t) & = & {C_P^W \over m_p} \sum_{i = 1}^N {\vec{u}_p(t) - \vec{u}_i(t) \over \|\vec{u}_i(t)-\vec{u}_p(t)\|^2}
 - {\nu_p \over m_p} \vec{v}_p(t). \\
  \dot{\vec{v}}_i(t) & = & {C_W^P \over m_i}{\vec{u}_p(t) - \vec{u}_i(t) \over \|\vec{u}_i(t)-\vec{u}_p(t)\|^2}\left(1 - {d_c^2(t) \over \|\vec{u}_i(t)-\vec{u}_p(t)\|^2} \right)\label{eq:3} \\
  & & \quad - \sum_{j = 1,j\ne i}^N {C_W^W \over m_i}{\vec{u}_j(t) - \vec{u}_i(t) \over \|\vec{u}_i(t)-\vec{u}_j(t)\|^2}\phi_{i,j}(t) - {\nu_i \over m_i} \vec{v}_i(t),\label{eq:4}
\end{eqnarray}
where $C_W^P$, $C_W^W$ and $C_P^W$ are positive constants adjusting the relative intensity of forces, so that \eg~the attraction that the prey exerts on the wolves is more intense than the repulsion than the wolves exert on the prey. This system must be solved with appropriate initial conditions, which should avoid pathological cases like when all agents are aligned. Parameter values shown in Table~\ref{table1} are those used in our previous work~\cite{Escobedo2014}.

\begin{table}
\begin{tabular}{|c|c|l|}
  \hline
  {\bf Parameter} & {\bf Value$^*$} & {\bf Physical meaning}\\ \hline
  $m_p$   & 1--2 & mass of the prey (elk: 350--400 Kg, bison: 700--900 Kg) \\
  $m_i$   &  0.1 & mass of wolf $W_i$ (35--40 Kg) \\
  $\nu_p$ & 2--4 & prey friction coefficient \\
  $\nu_i$ &   1  & wolf friction coefficient \\
  \hline 
  $C_W^P$ &  2  & coefficient of the force that the prey exerts on a wolf \Big. \\
  $C_W^W$ & 0.5 & coefficient of the interaction force between wolves \Big.\\
  $C_P^W$ & 0.2 & coefficient of the force that a wolf exerts on the prey \Big.\\
  \hline 
  $d_c$           & 1--2 & safe distance for wolves not to be harmed by the prey \\
  $d_a$           & 1--2 & avoidance distance for wolves to move away from each other \\
  $c_w$           & 0.5 & width coefficient of Gaussian function $\phi_{i,j}$ (width=$1/\sqrt{2c_w}$) \\
  \hline
  \end{tabular}
  \vspace{0.5em}
  \caption{\small\em Parameter values. $^*$Typical values taken from~\cite{Escobedo2014}.}
  \label{table1}
\end{table}

More details on the ethology of wolf-pack hunting strategies or on Canids behavior can be found in the original introduction of the model and in references therein~\cite{Muro2011, Escobedo2014}.


\subsection{Critical distances}

The parameter $d_c(t)$ denotes the {\it safe distance} at which a wolf stops to approach the prey in order to avoid to be armed during a possible counterattack of the prey. The role of $d_c(t)$ in Eq.~\eqref{eq:3} is to delimit the regions where the wolf is attracted or repulsed by the prey. Denoting by $R_i(t) = \|\vec{u}_i(t)-\vec{u}_p(t)\|$ the instantaneous distance of the $i$-th wolf to the prey, we have that, when the wolf is far from the prey, \ie, $R_i(t) > d_c(t)$, the force is attractive, while when the wolf is too close to the prey, i.e., $R_i(t) < d_c(t)$, the prey is repulsed. The value $d_c(t)$ is thus the balance point $R_i(t) = d_c(t)$ at which the wolf is not attracted nor repulsed by the prey.

When wolves approach the prey, a larger critical distance $d_a(t)>d_c(t)$ exists at which the wolves start to move away from each other, more specifically, from those other wolves that are also at this distance from the prey. This short-range repulsion is due to the natural need of individual space and to collision avoidance, and is stronger in stress situations as in the presence of a prey; a larger individual space is needed to have a better visibility of the prey and to move freely in response to possible attacks from the prey~\cite{Muro2011, Escobedo2014}. The effects of this second critical distance is introduced in the model through a repulsive interaction between wolves that becomes active when both wolves are at a distance $d_a(t)$ to the prey. This is the sum term in Expression~\eqref{eq:4}, whose value contributes to the behavior of the $i$-th and $j$-th wolves when the function $\phi(R_i(t),R_j(t))$ is not negligible:
\begin{eqnarray*}
  \phi_{i,j} = \phi(R_i(t),R_j(t)) = \exp\left\{-c_w \Big[ \big(R_i(t)-d_a(t)\big)^2 + \big(R_j(t)-d_a(t)\big)^2 \Big]\right\}.
\end{eqnarray*}
The function $\phi_{i,j}$ is a Gaussian function centered in $(d_a(t),d_a(t))$ and of width $1/\sqrt{2c_w}$. Its maximum value is one and is reached when both wolves $W_i$ and $W_j$ are at distance $d_a$ from the prey: $R_i(t) = R_j(t) = d_a(t)$. The intensity of the repulsion between $W_i$ and $W_j$ goes rapidly to zero as one of them is far from being at distance $d_a(t)$ from $P$.

The value and behavior of $d_c(t)$ and $d_a(t)$ are different for each wolf, depending mainly on the history of the wolf in previous hunting events, but also on the size and the health of the wolf. Here we will assume that wolves are homogeneous and have the same perception of the state of the prey, so that $d_c(t)$ and $d_a(t)$ preserve their respective value and time-variation across all wolves: $d_c^i(t) = d_c(t)$ and $d_a^i(t) = d_a(t)$ $\forall i=1,\dots,N$.

A key point for the analysis of how the optimal pack size depends on the size of the prey is that both critical distances $d_c(t)$ and $d_a(t)$ vary in time due to their dependence on the instantaneous perception of the state of the prey that wolves have. At the beginning of the hunt, the prey is fresh to react and even persecute and harass the wolf so that $d_c(t)$ and $d_a(t)$ are at their higher respective value. As the prey gets tired, the wolves become more confident and the critical distances decrease regularly, although not monotonically, because the prey is still able to display sudden reactions to try to injure some of the surrounding wolves, making $d_c(t)$ and $d_a(t)$ to increase abruptly.

More interestingly, $d_c(t)$ and $d_a(t)$ take larger values when hunting more dangerous prey. As pointed out by MacNulty {\it et~al.}~(2014), bison are the most difficult prey for wolves to kill, three times more difficult to kill than elk, with respect to which bison are not only larger, but also more aggressive and more likely to injure or kill wolves that attack them~\cite{MacNulty2014}.

\begin{figure}
\epsfxsize=0.89\textwidth
\begin{center}
 \epsfbox{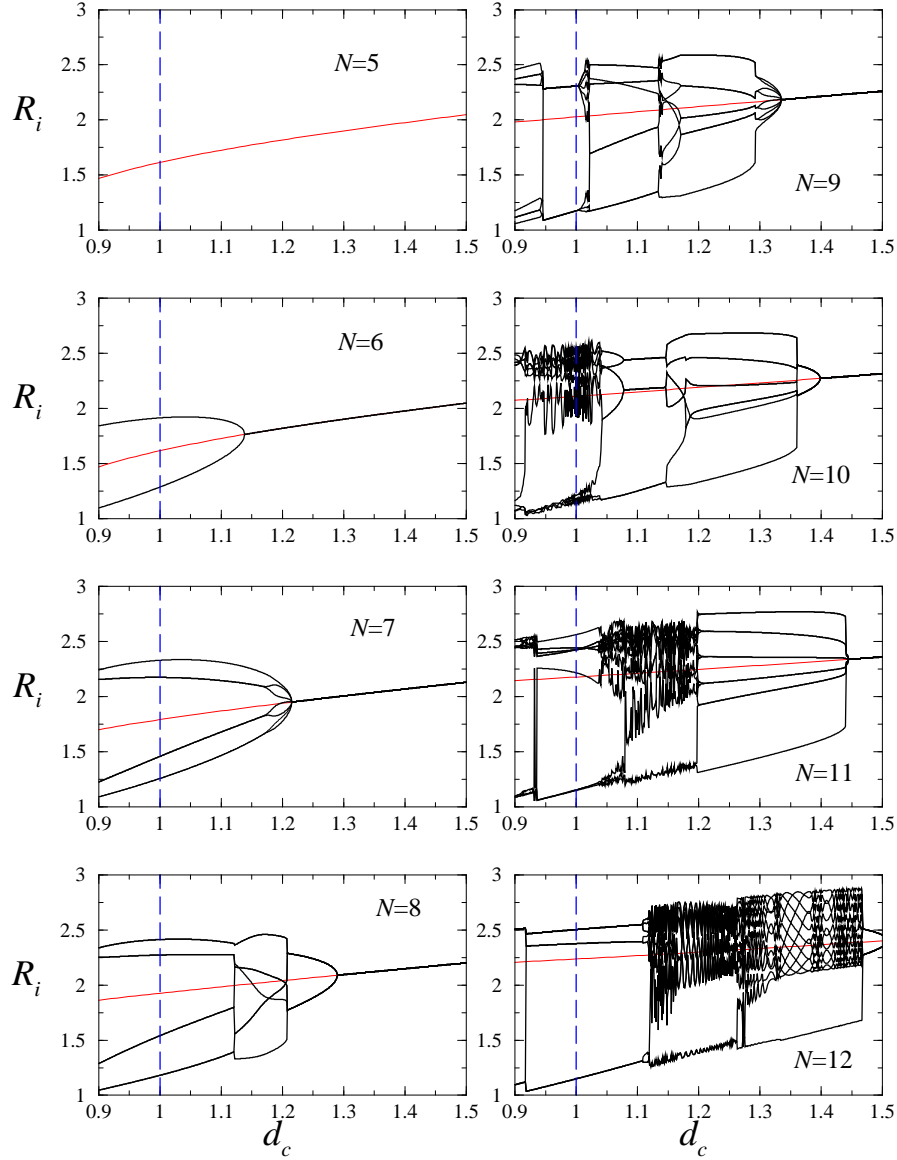}
\end{center}
\caption{\small\em Bifurcation diagram for different wolf-pack sizes when hunting elk (\ie, $d_a = 1.5$). Vertical axis: wolf-prey distance $R_i = \|\vec{x}_i - \vec{x}_p \|$ for $i = 1,\dots,N$; horizontal axis: $d_c \in [0.9,1.5]$, for $N=5,\dots,12$. Solid (black) lines denote the radius of the orbits where wolves are positioned when the stable spatial configuration is reached for the corresponding value of $d_c$. Multi-valued intervals correspond to multi-orbital configurations. The bifurcation value $d_c^*$ separates single-valued intervals from multivalued ones. For $N = 5$ there is no multi-valued intervals, for $N = 6$, $d_c^* \approx 1.14$. Dashed (red) line denotes the radius of the regular polygon (color online).}
\label{fig3}
\end{figure}

The variation of the spatial configuration that the wolf pack exhibits along a hunting exercise is as follows. At the beginning of the hunt, $d_c(t)$ is large and the wolf-pack tends to a spatial configuration described by a regular polygon (RP) of $N$ vertices and radius $R_{\rm RP}^N(d_c(t))$. If $d_c(t)$ varies slowly, the regular polygon is almost stationary. When $d_c$ is above $d_c^*$, the stationary regular polygon (SRP) is stable. As the prey gets tired, $d_c(t)$ decreases and can cross the bifurcation point $d_c^*$ (provided $d_c^* > d_s$), so that $d_c(t)$ is in the Grey region, where the SRP is unstable: a small perturbation of the regular polygon will disrupt the formation and the wolves will split into two or more orbits, with (at least) one orbit closer and (at least) one orbit further from the prey than $R_{\rm RP}^N$, that is, $R_{\rm in} < R_{\rm RP} < R_{\rm out}$, where $R_{\rm in}$ (resp.~$R_{\rm out}$) is the radius of the closest (resp.~furthest) orbit to the prey.

Which orbit is a privileged position for wolves to stay depends on the behavior of the prey: if the prey is at the end of the struggle, wolves in the inner orbit have a better chance to approach the prey and start to dissect it, while if the prey is still able to display a sudden counterattack, privileged positions are those in the outer orbit, where wolves have a better chance to avoid the blow and escape.

For example, a pack of 7 wolves will converge towards a regular heptagon of radius $R_{\rm RP}^N = 1.97$ when $d_c(t) = 1.25$, but, if another wolf joins the hunt, the heptagon is destabilized and privileged positions emerge. For $N = 8$ and $d_c = 1.25$, the stable spatial configuration is a two-orbits configuration with radii $R_1 = 1.88$ and $R_2 = 2.24$, while the radius of the SRP is $R_{\rm RP} = 2.07$; see Fig.~\ref{fig2}.

Similarly, when the prey of a wolf-pack of $7$ wolves is getting tired, the value of $d_c(t)$ decreases below $d_c^*$ and the regular polygonal configuration becomes unstable in benefit of the multi-orbital configuration, leading again to complex behavioral patterns with the emergence of privileged positions and enhancing the possibility of disrupting the hunt. See the numerical simulations in the Supplementary material of~\cite{Escobedo2014}.

Fig.~\ref{fig3} shows the bifurcation point $d_c^*$ for a given value of $d_a = 1.5$ and wolf-pack sizes from $N = 5$ to $12$. These diagrams have been obtained by solving numerically the wolf-prey system for a fixed value of $d_c$ until a stable equilibrium is reached. Stable solutions can be stationary or not (see again~\cite{Escobedo2014}). From these diagrams, the value of $d_c^*$ is calculated as a function of $N$ and $d_a$, producing Fig.~\ref{fig1} (for different values of $d_a$) and Fig~\ref{fig2} (for $d_a = 1.5$).


\section*{Author Summary}

Social foraging shows unexpected features such as the existence of an optimal group size above which additional individuals do not favor the success of the hunt. Previous work shows that the optimal group size is surprisingly small. In wolves hunting elk in Yellowstone Park, hunting success levels off beyond pack sizes of 4 individuals. This observation recently received support from a computational agent model which showed that the reduction of hunting success in large packs can be due to the emergence of privileged positions in the spatial wolf-pack formation. Subsequent observations of wolves hunting bison reinforce and document the hypothesis of the privileged positions. When hunting bison, the optimal wolf-pack size is between 9 and 13. We show here that this is in accordance with the computational model. Moreover, although the optimal group size is expected to be greater when hunting more dangerous prey, we show that this relation is surprisingly not linear: the computational model reveals that the optimal group size actually results from the opposite contributions of two critical distances separating wolves and prey. These distances strongly depend on the kind of prey, and can induce a different variation if a different prey is considered ({\it e.g.}~moose).

\backmatter
\bibliography{biblio}
\bibliographystyle{smfplain}

\end{document}